\documentclass[conference]{IEEEtran}

\usepackage{newtxtext}

\newcommand{\muse}{\textit{\textsc{MuSe}}}

\usepackage{hyperref}
\usepackage{cleveref}
\usepackage[table]{xcolor}
\usepackage{graphicx}
\usepackage{booktabs}
\usepackage{balance}

\begin{document}
\title{MuSe: a Mutation Testing Plugin for the Remix IDE}




\author{
\IEEEauthorblockN{Gerardo Iuliano}
\IEEEauthorblockA{University of Salerno, Italy\\
geiuliano@unisa.it}

\and
\IEEEauthorblockN{Daniele Carangelo}
\IEEEauthorblockA{University of Salerno, Italy\\
d.carangelo@studenti.unisa.it}

\and
\IEEEauthorblockN{Carmine Calabrese}
\IEEEauthorblockA{University of Salerno, Italy\\
c.calabrese31@studenti.unisa.it}

\and
\IEEEauthorblockN{Dario Di Nucci}
\IEEEauthorblockA{University of Salerno, Italy\\
ddinucci@unisa.it}
}

\IEEEpeerreviewmaketitle
\maketitle


\begin{abstract}
Mutation testing is a technique to assess the effectiveness of test suites by introducing artificial faults into programs. Although mutation testing plugins are available for many platforms and languages, none is currently available for \textsc{Remix-IDE}, the most widely used Integrated Development Environment for the entire contract development journey, used by users of all knowledge levels, and serves as a learning lab for teaching and experimenting with Ethereum. The quality and security of smart contracts are crucial in blockchain systems, as even minor issues can result in substantial financial losses. This paper proposes \muse{}, a mutation testing plugin for the \textsc{Remix-IDE}. \muse{} includes traditional, Solidity-specific, and security-oriented mutation operators. Its integration into the \textsc{Remix-IDE} eliminates the need for additional setup and lowers the entry barrier. As a result, developers and researchers can immediately leverage mutation testing to assess the effectiveness of their test suites and identify potential issues in smart contracts. 
We provide a demo video showing \muse{}\footnote{\url{https://www.youtube.com/watch?v=MIFk9exTDu0}} and its repository\footnote{\url{https://github.com/GerardoIuliano/MuSe-Remix-Plugin}}.
\end{abstract}


\section{Introduction}
\label{sec:introduction}

In recent years, the interaction with decentralized applications (dApps) has increased. A considerable number of dApps exhibit predominantly positive usage trends across various categories, including decentralized finance, healthcare, exchanges, Gambling, games, marketplaces, and social media~\cite{bartl2023statistical}. Decentralized Applications are composed of smart contracts, self-executing programs that run on a blockchain. Smart contracts automatically enforce and execute the terms of an agreement when predefined conditions are met, without the need for intermediaries. They are developed in Integrated Development Environments (IDEs), which streamline the writing, testing, and deployment of smart contracts by providing features such as code editing, debugging, and automated testing. The latter is essential, given the immutable nature of smart contracts once deployed on a blockchain.

The adoption of IDEs is growing as the smart contract market expands, driven by the promise of reduced costs, increased efficiency, and automation across industries. However, according to Zou et al.~\cite{testing_challenge}, significant challenges remain in adopting smart contracts. On the one hand, no mature testing frameworks comparable to those in other programming languages are available, making it challenging to cover all corner cases and scenarios. On the other hand, testing can be costly if conducted on \textsc{testnets} or \textsc{mainnet}, the public blockchain networks used respectively for deployment testing and real-world execution..
Additionally, a few tools measure the quality of a smart contract test suite~\cite{testing_challenge}.

Mutation testing is typically used to evaluate the adequacy of test suites, to guide test case generation, and to support experimentation~\cite{papadakis2019mutation}. In recent years, some tools have been released, such as \textsc{Deviant}~\cite{deviant}, \textsc{RegularMutator}~\cite{regularmutator}, 
\textsc{Musc}~\cite{musc}, and \textsc{SuMo}~\cite{sumo}. 
Nevertheless, none of these tools is integrated into an IDE, which hinders their use.

This paper describes \muse{}, our mutation testing tool based on \textsc{SuMo}, which is the most complete tool in the state of the art. Both tools rely on \textsc{Solidity-parser-antlr}, a parser generated from a robust ANTLR4 grammar that produces an Abstract Syntax Tree of the code. 
\muse{} implements six new security-oriented mutation operators that can be used to verify whether the testing codebase could identify such security issues or to generate synthetic datasets of vulnerable code.
Our mutation operators leverage \textsc{Solidity-parser-antlr} to identify injection patterns where the vulnerabilities can be injected. 
The tool is wrapped into a \textsc{Remix-IDE} plugin to facilitate its use during smart contract development and testing directly within the IDE. We selected \textsc{Remix-IDE} because it is the most widely used IDE to develop Solidity smart contracts. It requires no setup, fosters a fast development cycle, and has a rich set of plugins with intuitive GUIs. The IDE is available as a web or desktop app. To the best of our knowledge, the \muse{} plugin for \textsc{Remix-IDE} is the first mutation testing tool integrated into the IDE.
We evaluated the injected vulnerabilities in a previous work~\cite{ease25} by verifying whether Slither~\cite{slither}, a state-of-the-art static analysis tool for smart contracts, can identify them. The results reveal that each vulnerability has a distinct injection rate, and successful injection depends on whether the smart contract satisfies the preconditions of the vulnerability pattern. 


The paper is structured as follows. \Cref{sec:related_work} reviews the background and related literature. \Cref{sec:muse} presents the \muse{} plugin and its mutation operators, and \Cref{sec:howto} explains how to install and use the plugin in \textsc{Remix}. \Cref{sec:evaluation} reports the experimental results, while \Cref{sec:impact} discusses the potential impact. Finally, \Cref{sec:conclusion} concludes the paper.


\section{Background and Related Work}
\label{sec:related_work}

Smart contracts were introduced by Nick Szabo in 1994 and have found new life on the Ethereum blockchain. Smart contracts on Ethereum are self-executing contracts whose terms are directly written in code~\cite{antonopoulos2018masteringSc}. They automatically execute predefined conditions, ensuring trust and reliability. These contracts run on the Ethereum network, a decentralized platform that eliminates the need for intermediaries by automating contract execution. Each smart contract has a unique address and is initiated by a transaction between a sender and the contract. Execution incurs a computational cost measured in gas, which regulates resource usage and rewards miners. Once deployed on the blockchain, these contracts become immutable and tamper-proof~\cite{kaushal2021immutable}, making their security critical.

Given the essential importance of ensuring smart contract security, mutation testing offers a systematic approach to assessing the robustness of testing used to detect potential faults.
Mutation testing is a software testing technique to evaluate the effectiveness of test cases by intentionally introducing small changes, called mutants, into the source code to simulate potential faults or errors. The primary goal is to assess whether the existing test cases can detect these changes, thereby measuring the fault-detection capability of a test suite~\cite{PAPADAKIS, mutationTesting}. This process ensures software is rigorously validated, improving its reliability and reducing the likelihood of undetected faults in production.

The literature provides several tools to support mutation testing for Solidity smart contracts.

Chapman et al. introduced \textsc{Deviant}~\cite{deviant}, a mutation testing framework that automatically generates mutated versions of Solidity projects and evaluates them against existing test suites. The tool incorporates both Solidity-specific mutation operators, derived from a dedicated fault taxonomy, and traditional programming operators to simulate faults. The authors applied Deviant to three Solidity projects, demonstrating that high statement and branch coverage does not necessarily ensure strong test quality. Their findings highlight the importance of rigorous testing practices in mitigating financial risks.

Ivanova and Khritankov developed \textsc{RegularMutator}~\cite{regularmutator}, a mutation testing tool that leverages regular expressions to inject Solidity-specific faults. The mutation operators implemented in \textsc{RegularMutator} reflect common developer mistakes, enabling the simulation of realistic faults. An empirical evaluation of large-scale Solidity projects revealed that mutation analysis provides a more reliable assessment of test suite quality than traditional coverage metrics.

Li et al. proposed \textsc{MuSC}~\cite{musc}, a robust and user-friendly mutation testing tool for Ethereum smart contracts. It performs mutation operations at the Abstract Syntax Tree (AST) level, ensuring both precision and efficiency in fault injection. Furthermore, \textsc{MuSC} supports user-defined testnet configurations, enabling developers to meet diverse needs. It implements a comprehensive set of Solidity-specific mutations.

Barboni et al. proposed \textsc{SuMo}~\cite{sumo}, a mutation testing tool that incorporates 25 Solidity-specific operators and 19 general-purpose operators. The tool enables systematic evaluation of test suite effectiveness in Solidity projects. 
 \textsc{SuMo} was later extended into \textsc{ReSuMo}~\cite{resumo}, which introduces a regression mutation testing approach. It relies only on the Truffle testing framework. \textsc{ReSuMo} applies a static, file-level strategy to selectively mutate only relevant smart contracts and rerun corresponding test cases during regression testing. By incrementally updating results based on prior test outcomes, \textsc{ReSuMo} improves efficiency while reducing computations.

Our tool covers the limitations of state-of-the-art tools.
\textsc{Deviant} focuses on traditional mutation and Solidity-specific mutations. It relies on the \textsc{solc} compiler 0.4.24, making it compatible only with Solidity versions from 0.4.0 to 0.4.24. Additionally, it is not open-source.
\textsc{RegularMutator} implements Solidity-specific operators, focusing on vulnerable mutation to inject three different Solidity vulnerabilities. Unfortunately, it is not open source.
\textsc{Musc}, \textsc{SuMo}, and \textsc{ReSuMo} implement traditional mutation operators and Solidity-specific operators. They are open-source and accept Truffle projects. Unfortunately, they do not implement security-oriented operators and cannot be used directly in \textsc{Remix-IDE}.
\muse{} is built on top of \textsc{SuMo}, adding six new security-oriented mutation operators. It provides the most comprehensive set of mutation operators, supports Solidity versions from 0.4.x to 0.8.x, and can be accessed directly within \textsc{Remix-IDE}.
\section{The \muse{} Plugin}
\label{sec:muse}

\muse{} extends \textsc{SuMo}~\cite{sumo}, which is implemented in JavaScript/TypeScript and distributed as an NPM package. It leverages the \textsc{solidity-parser-antlr} library to parse Solidity source code into an Abstract Syntax Tree (AST).
All mutation operations are at the AST level and leverage the Visitor design pattern, ensuring syntactically correct mutants and avoiding errors introduced by naive text-based approaches.

\begin{table*}[ht]
    \centering
    \caption{Vulnerabilities injected into smart contracts using the security-oriented mutation operator of \muse{}.}
    \rowcolors{2}{gray!10}{white}
    \resizebox{\textwidth}{!}{
    \begin{tabular}{l p{13cm} c}
        \toprule 
        \textbf{Vulnerability} & \textbf{Description} & \textbf{Operator}\\ 
        \midrule
        Unchecked low-level call return value & Low-level calls return \textit{false} on failure instead of throwing exceptions, risking critical vulnerabilities if unchecked.  & UC\\
        Unchecked send & The send function returns \textit{false} on failure without throwing an exception, risking vulnerabilities if unchecked. & US\\
        Authentication via tx.origin & Using \textit{tx.origin} for authorization risks vulnerabilities if an authorized account interacts with a malicious contract. & TX\\
        Unused return & The return value of an external call is not stored in a local or state variable. & UR\\
        Multiple calls in a loop & Calls inside a loop might lead to a denial-of-service attack. & CL\\ 
        Delegatecall to untrusted callee & \textit{Delegatecall} executes the code at the target address in the context of the calling contract. It allows a SC to load code dynamically from a different address. & DTU\\ 
        \bottomrule
    \end{tabular}
    }
    \label{tab:vuls}
\end{table*}

\subsection{\muse{} Mutation Approach}
The mutation approach aims to make mutation testing practical and effective for Solidity smart contracts by addressing key challenges in fault modeling, usability, and efficiency. Its design begins with a comprehensive fault model, derived from an in-depth analysis of the Solidity language and existing vulnerability taxonomies~\cite{iulianoSLR}, to define meaningful mutation operators that simulate realistic programming mistakes and vulnerability patterns.
\muse{} uses a customized mutation process that allows testers to selectively apply mutation operators and target specific contracts, thereby reducing computational overhead while maintaining effectiveness. To further optimize performance, \textsc{SuMo} tackles the generation of stillborn mutants, those trivially rejected by the compiler, by excluding ineffective operators and introducing precondition checks based on the contract’s AST, ensuring that mutations produce syntactically valid code.
To minimize redundancy, \muse{} employs two strategies for reducing subsumed mutants: (1) Operator Optimizations, which merge or simplify mutation rules likely to generate equivalent test outcomes, and (2) Trivial Compiler Equivalence (TCE), which uses compiler optimization techniques to identify and discard functionally equivalent mutants automatically.
Additionally, \muse{} integrates mechanisms to remove equivalent mutants that cannot be detected by any test suite, leveraging TCE to automatically eliminate up to 30\% of them. Overall, the framework balances completeness with efficiency, enabling a more scalable and meaningful evaluation of test suite effectiveness in detecting faults and vulnerabilities in smart contracts.
Finally, \muse{} integrates with mainstream smart contract development frameworks, including \textit{Truffle}, \textit{Hardhat}, \textit{Brownie}, and \textit{Foundry}.
The integration is enabled by configuration files and CLI commands, which will allow developers to select mutation operators to apply, specify contracts/files to mutate, select random sampling of mutants, set a timeout threshold, and define the inclusion/exclusion of test cases.
\muse{} is released under the GPL-3.0 licence.

\subsection{\muse{} Mutation Operators}
\textsc{SuMo} already provides 25 Solidity-specific and 19 general-purpose mutation operators, many of which extend or refine existing rules. We added six new security-oriented operators, listed in \Cref{tab:vuls}, to introduce faults into realistic scenarios, enabling the generation of contracts with vulnerabilities placed in both typical and unconventional yet valid locations, and challenging the test suite to expand its scope to security testing.
Overall, \muse{} comprises 50 mutation operators:

\begin{itemize}
    \item \textbf{25 Solidity-specific operators} derived from common programming mistakes and Solidity language peculiarity. Examples include modifying function visibility, removing the payable modifier, altering self-destruct or transfer instructions, and changing Ether value transfer conditions.
    
    \item \textbf{19 traditional programming operators} including arithmetic, relational, and logical operators, as well as literal modification and boolean negation.

    \item \textbf{6 security-oriented operators} allowing the injection of the following vulnerabilities: Unchecked low-level call return value, Unchecked send, Authentication via tx.origin, Unused return, Multiple calls in a loop, and Delegatecall to untrusted callee~\cite{iulianoSLR, ramederSLR, openscv}.
\end{itemize}

    


\section{How to Use \muse{}}
\label{sec:howto}

In this paper, we provide \muse{} as a plugin for \textsc{Remix IDE}, whose Plugin Manager allows integrating tools developed by the Remix team with those created by external teams. 

\begin{figure}[htbp]
    \centering
    \includegraphics[width=0.99\columnwidth]{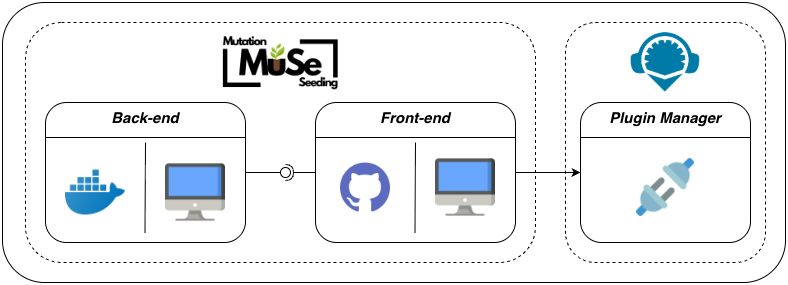}
    \caption{Architecture of the \muse{} plugin}
    \label{fig:muse_architecture}
\end{figure}

\begin{figure}[htbp]
    \centering
    \includegraphics[width=0.85\columnwidth]{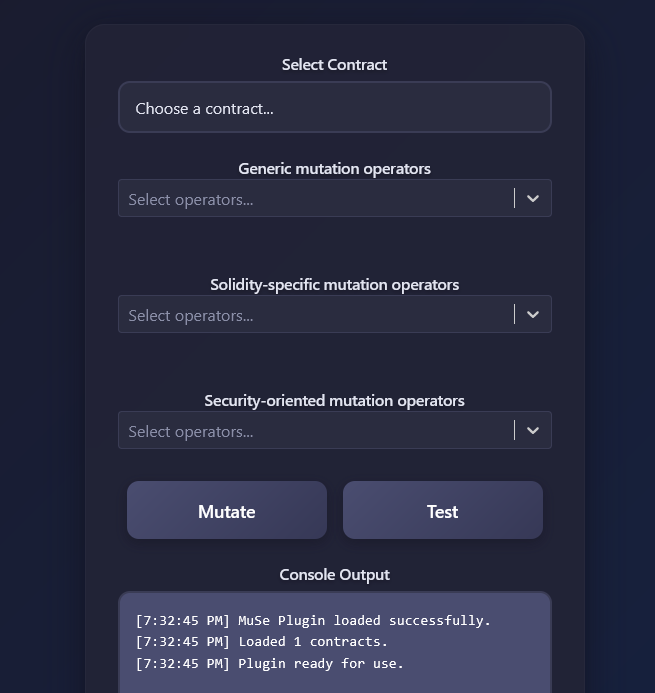}
    \caption{\muse{} user interface in \textsc{Remix-IDE}}
    \label{fig:muse_ui}
\end{figure}

To use \muse{} as a plugin for \textsc{Remix IDE}, developers need to install \muse{} locally. First, they need to download the tool from our GitHub repository$^2$. The recommended installation option is to pull and run the \muse{} Docker image from \textsc{DockerHub}. The alternative is to clone the tool from GitHub and set it up locally following the installation guide$^2$.
As shown in \Cref{fig:muse_architecture}, \muse{} has two modules: one for the back-end and another for the front-end. These can be installed locally by cloning the GitHub repository and setting them up. For a local setup, \textsc{Node.js} version 20.19.0 or higher is required. 
After installing the tool, developers need to connect it to \textsc{Remix IDE} as a Local Plugin by selecting ``Connect to a Local Plugin'' in the IDE Plugin Manager and entering the URL to connect the plugin front-end to the IDE. Once the plugin is installed and connected, the \muse{} Plugin tab appears in the \textsc{Remix IDE} sidebar as shown in the \Cref{fig:muse_ui}.

\begin{figure}[htbp]
    \centering
    \includegraphics[width=0.9\columnwidth]{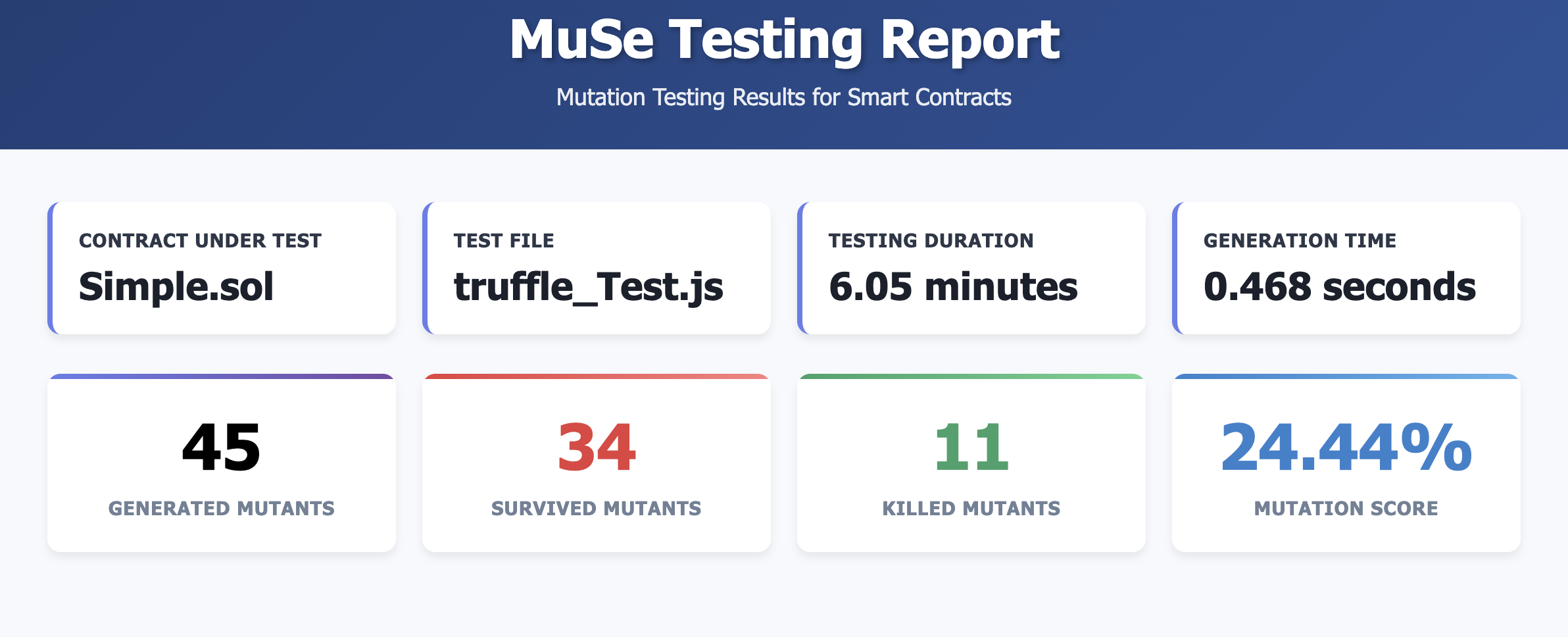}
    \caption{\muse{} HTML report}
    \label{fig:muse_report}
\end{figure}

The \muse{} Plugin User Interface enables selecting the contract to mutate among all compiled contracts in \textsc{Remix IDE} under the contracts folder.
The developer can choose which mutations to inject using three selectors, one for each mutation category listed before. Finally, the ``Mutate'' button starts the mutation process.
After these steps, all logs are displayed in real-time in the console. The console also shows the number of generated mutants, which are automatically added to the \textsc{Remix} workspace under the \muse{} folder.
At this point, developers can click the ``Test'' button to configure the test framework by selecting one of the four available options (Truffle, Hardhat, Brownie, Forge), and setting the timeout in seconds to end the test. Once the testing phase is completed, the console displays a summary of the results. \muse{} automatically generates a HTML report under the \muse{} folder of the project. The report provides an overview of the results, with an summary of executed tests and their outcomes. \Cref{fig:muse_report} shows an example of a report including generated, survived, and killed mutants, and the final mutation score.

We provide a demo video showing \muse{}$^1$, while the installation guide is available in our online appendix$^2$.
\section{\muse{} Evaluation}
\label{sec:evaluation}

\muse{} extends SuMo~\cite{sumo}, implementing a new set of security-oriented mutation operators.
Therefore, Solidity-specific operators and traditional programming operators were not re-validated, as they were analyzed in previous work~\cite{sumo}. 
To ensure the correctness of our security-oriented mutation operators, we manually validated a statistically significant set of generated smart contracts extracted from the SmartBugs-Wild dataset~\cite{smartbugs_wild}, which contains 47,398 smart contracts from the Ethereum blockchain. \muse{} generated  350,716 mutated contracts. Using a 95\% confidence level and a 5\% margin of error, we determined a statistical sample size of 384 contracts, randomly selected for manual inspection.
The validation procedure consisted of four steps:


    

\textit{\textbf{Syntactic Validation.}} Each contract in the sample was compiled to verify syntactical correctness.

\textit{\textbf{Log Comparison.}} We compared the mutation logs produced by SuMo, which specify the mutation type and the affected lines of code, against the corresponding mutated contracts.

\textit{\textbf{Pattern Verification.}} We checked whether the injected mutations adhered to the patterns defined by the corresponding operator, ensuring that vulnerabilities were introduced at the correct program locations.

\textit{\textbf{Modification Assessment.}} We inspected the modifications to confirm that the mutations either (i) inserted new code introducing the target vulnerability or (ii) altered existing code to make the contract vulnerable to the intended issue.

A mutation was considered valid if it satisfied either of the conditions in step 4. The results show that \muse{} failed to inject vulnerabilities in 20 out of 384 cases (5.21\%). These failures were exceptional cases that the mutation operators did not fully address. We refined the mutation operators based on the results of this manual inspection.

\paragraph{Injection rate of each injected vulnerability}

Starting from the 350,493 vulnerable mutants generated by \muse{} on the SmartBugs-Wild dataset, we examined the injection rate of the six security-oriented mutation operators.

\begin{table}[h]
    \centering
    \caption{Results for mutation operators ordered by injection rate, total number of generated mutants, and average injection rate of \muse{}}
    \label{tab:results_rq1}
    \rowcolors{2}{gray!10}{white}
    \begin{tabular}{lrrr}
    \toprule
        \textbf{Operator} & \textbf{\# Mutated SCs} & \textbf{\# Mutants} & \textbf{Injection Rate}\\ \midrule
        UR  & 33,910 & 213,912 & 71.50\%\\
        TX  & 32,250 & 65,825 & 68.00\%\\
        CL  & 26,604 & 61,687 & 56.00\%\\
        UC  & 4,094 & 4,992 & 8.60\%\\
        US  & 2,248 & 3,928 & 4.70\%\\
        DTU & 113 & 149 & 0.23\%\\ \midrule
        -  &  -  & 350,493 & 34.83\%\\
        \bottomrule
    \end{tabular}
\end{table}

As shown in \Cref{tab:results_rq1}, the most prevalent injectable vulnerability is the unused-return pattern (UR), with a 71.5\% injection rate. The patterns used to inject vulnerabilities are not only common but also frequently appear within contracts, with an average of six occurrences per contract, making them natural targets for mutation. The second most injectable vulnerability is authorization via tx.origin (TX), which is injectable in 68\% of cases (the TX operator performs mutations when ownership or privilege checks involve msg.sender). Multiple calls in a loop (CL) have an injection rate of 56\%, indicating that just over half of contracts use call/send/transfer; 8\% of contracts already contained the target vulnerability before mutation. Unchecked low-level call (UC) and unchecked send (US) follow with injection rates of 8.6\% and 4.7\%, respectively, and the call primitive appears nearly twice as often as send; call is more flexible but carries greater reentrancy risk. Delegatecall-to-untrusted-callee (DTU) has the lowest injection rate, reflecting its rarer, specialised use (e.g., proxy patterns). Because some contracts lack the specific patterns, constructs, or Solidity functions required for particular injections, not every vulnerability can be injected into every contract. Overall, applying the mutation operators expanded the vulnerable dataset by approximately 840\%.

\paragraph{Detection rate of each injected vulnerability}

\begin{table}[ht]
    \centering
    \caption{Detection rate of injected vulnerability and overall performance of Slither on the six considered vulnerabilities}
    \label{tab:results_rq2}
    \rowcolors{2}{gray!10}{white}
    \begin{tabular}{lrrrr}
    \toprule
        \textbf{Vulnerability} & \textbf{TP} & \textbf{FN} & \textbf{Recall} & \textbf{FNR}\\ \midrule
        UC & 4,876 & 0 & 1.000 & 0.000\\
        US & 3,570 & 0 & 1.000 & 0.000\\
        CL & 45,261 & 10,563 & 0.810 & 0.189\\
        UR & 124,858 & 81,184 & 0.605 & 0.394\\
        TX & 21,765 & 42,937 & 0.336 & 0.663\\
        DTU & 15 & 134 & 0.100 & 0.899\\ \midrule
        -   & 200,345 & 134,818 & 0.597 & 0.402\\
        \bottomrule
    \end{tabular}
\end{table}

We ran Slither on the mutants generated by \muse{} to evaluate its ability to detect injected vulnerabilities. Assuming correct injection, the detection results before and after the mutation were compared. As shown in \cref{tab:results_rq2}, Slither achieved perfect detection (recall = 1.00) for unchecked low-level call return value (UC) and unchecked send (US), demonstrating strong detectors for verifying proper handling of return values. Detection was also high for multiple calls in a loop (CL) with a recall of 0.81, though some vulnerabilities remained undetected. Performance dropped for unused returns (UR) with a recall of 0.63, and was notably weak for authorization via tx.origin (TX) and delegatecall to untrusted callee (DTU), which reached recall of 0.33 and 0.10, respectively. Overall, Slither achieved an average recall of 0.597, indicating that although it effectively identifies specific vulnerabilities, it misses approximately 40\% of injected cases. These underscore the need for improved detection mechanisms. At the same time, the results show that injected vulnerabilities are also helpful to assess the effectiveness of a test suite.
\section{\muse{} Potential Impact}
\label{sec:impact}
The integration of \muse{} directly into \textsc{Remix-IDE} significantly advances smart contract testing and security.

By embedding mutation testing into the development workflow, developers gain immediate feedback on the effectiveness of their test suites, facilitating validation and promoting a more proactive approach to fault detection and prevention. As developers continuously interact with mutants within the IDE, they can iteratively refine and strengthen their test cases, ultimately improving code robustness and reliability. \muse{} enables a culture of continuous testing, reduces the likelihood of undetected vulnerabilities, and enhances overall software quality. In the context of smart contract development, where immutability and security are essential, an IDE-integrated mutation testing tool can significantly improve the security and trustworthiness of blockchain applications.

\muse{} can also be used as a benchmark generator~\cite{ease25}. Using its security-oriented mutation operators, it is possible to inject vulnerabilities in smart contracts and use them to evaluate the performance of vulnerability detection tools. 

\section{Conclusion}
\label{sec:conclusion}

This paper presents \muse{}, a mutation testing plugin for \textsc{Remix-IDE}. The tool features 50 mutation operators targeting Solidity-specific features, general programming constructs, and security-oriented patterns. \muse{} allows developers to evaluate the effectiveness of their test suite in detecting faults and vulnerabilities. We have already preliminarily assessed the ability of the tool to inject vulnerabilities. As part of our future work, we plan to implement additional security-oriented mutation operators, validate them, and empirically evaluate test suite effectiveness using \muse{}.

\section*{Acknowledgments}
Funded by the European Union – Next Generation EU, Mission 4 Component 1, CUP D53D23008400006, and by the Italian Government (MUR, PRIN 2022) code 202233YFCJ. Gerardo Iuliano is funded by the European Union – Next Generation EU, Mission 4 Component 2, CUP D42B24002220004, by the Italian Government (MUR), and by AstraKode S.r.l.

\bibliographystyle{IEEEtran}
\bibliography{IEEEabrv}

\balance

\end{document}